\documentclass[12pt]{article}
\usepackage{graphicx}
\usepackage{amssymb}
\usepackage{amsmath}

\usepackage{bm}

\newcommand{\rf}[1]{(\ref{#1})}
\newcommand{\beq}{\begin{equation}}

\newcommand{\eeq}{\end{equation}}
\newcommand{\bea}{\begin{eqnarray}}
\newcommand{\eea}{\end{eqnarray}}

\newcommand{\e}{\mbox{e}}

\newcommand{\La}{\Lambda}

\renewcommand{\a}{\alpha}

\newcommand{\eps}{\epsilon}

\newcommand{\del}{\delta}
\newcommand{\Del}{\Delta}

\renewcommand{\k}{\kappa}

\newcommand{\ra}{\rangle}
\newcommand{\la}{\langle}

\begin{document}

\begin{center}
\vspace{24pt}
{ \Large \bf The Semiclassical Limit of Causal Dynamical}
\vspace{6pt}\\
{\Large \bf Triangulations}

\vspace{30pt}

{\sl J. Ambj\o rn}$\,^{a,c}$,
{\sl A. G\"{o}rlich}$\,^{b}$,
{\sl J. Jurkiewicz}$\,^{b}$,
{\sl R. Loll}$\,^{c,d}$,
{\sl J. Gizbert-Studnicki}$\,^{b}$
and {\sl T. Trze\'{s}niewski}$\,^{b}$

\vspace{24pt}
{\footnotesize

$^a$~The Niels Bohr Institute, Copenhagen University\\
Blegdamsvej 17, DK-2100 Copenhagen \O , Denmark.\\
{ email: ambjorn@nbi.dk}\\

\vspace{10pt}

$^b$~Institute of Physics, Jagellonian University,\\
Reymonta 4, PL 30-059 Krakow, Poland.\\
{ email: atg@th.if.uj.edu.pl, jurkiewicz@th.if.uj.edu.pl,
jakub.gizbert-studnicki@uj.edu.pl, t.trzesniewski@uj.edu.pl}\\

\vspace{10pt}

$^c$~Institute for Theoretical Physics, Utrecht University, \\
Leuvenlaan 4, NL-3584 CE Utrecht, The Netherlands.\\
{ email: r.loll@uu.nl}\\

\vspace{10pt}

$^d$~Perimeter Institute for Theoretical Physics, \\
31 Caroline St. N., Waterloo, Ontario, Canada N2L 2Y5.\\ 

\vspace{10pt}
}
\vspace{48pt}

\end{center}


\begin{center}
{\bf Abstract}
\end{center}
Previous work has shown that the macroscopic structure of 
the theory of quantum gravity defined by causal dynamical triangulations
(CDT) is compatible with that of a de Sitter universe. 
After emphasizing the strictly nonperturbative nature of this 
semiclassical limit we present a detailed study of the three-volume data,
which allows us to re-confirm the de Sitter structure,
exhibit short-distance discretization effects, and make a first detailed 
investigation of the presence of higher-order curvature terms in  
the effective action for the scale factor.
Technically, we make use of a novel way of fixing the total four-volume
in the simulations.

\newpage

\section{Introduction}\label{intro}

Mundane field-theoretical descriptions of quantum gravity 
have been undergoing a renaissance, 
thanks to the use of improved renormalization
group techniques \cite{RG}\footnote{building on the original idea 
of ``asymptotic safety" \cite{weinberg}},
new ideas about the relation between space and time at 
short scales \cite{horava}, and 
a novel way of implementing a lattice regularization of 
four-dimensional quantum gravity, which takes into account 
both the dynamical and the causal, Lorentzian nature of spacetime
\cite{ajld4}.\footnote{building on the earlier idea of ``dynamical
triangulation" as a regularization of quantum gravity
\cite{aj}, but incorporating a concept of micro-causality along the lines
of \cite{teitelboim}} 
New (and partially 
overlapping \cite{spectral,spectral-RG,spectral-hl,horava-simu,horava-new}) 
results obtained in
these approaches give rise to the hope that standard tools from quantum
field theory -- adapted to the situation where spacetime itself is dynamical --
are indeed sufficient to construct a theory of quantum gravity nonperturbatively.
This is an attractive prospect, since it may imply a large degree of uniqueness,
with only a small number of parameters needing to be fine-tuned to get to the
correct theory.

In this article we will discuss several 
aspects of the semiclassical limit of the research program on 
``Quantum Gravity from Causal Dynamical Triangulations (CDT)", 
in which a dynamical lattice provides
a geometric UV cut-off. In earlier papers we have reported that the 
infrared limit of CDT allows an interpretation as the classical 
solution to Euclidean Einstein gravity with a positive cosmological constant
\cite{emerge,blp,semi,agjl,bigs4} 
(see also \cite{contemp} for a pedagogical review).
Here we investigate this limit in more detail, including discussions
of discretization effects and of the asymmetry  
between space and time, which appears at the lattice-regularized stage
of the CDT set-up. 

It is important to bear in mind that what we call the semiclassical limit is 
a {\it truly nonperturbative} limit. This means that the tentative continuum limit 
of the path integral 
is to be found in a region of the bare coupling constant space 
where the entropy of various geometric configurations makes a contribution
which is {\it as} important as the contribution coming from the 
exponential of the action. In lower dimensions, this situation is illustrated by
the famous Kosterlitz-Thouless transition in the XY model.
The XY model is a lattice spin model, whose ``spins" are
two-dimensional vectors
of unit length. In two spatial dimensions, this model has vortex configurations, 
with an energy per vortex of approximately
\beq\label{0.0}
E= \k \ln (R/a),
\eeq
where $\k$ is a coupling constant, $R$ a measure of the linear size of the 
system and $a$ the lattice spacing. Ignoring boundary effects, the 
centre of the vortex can be placed at any one of the $(R/a)^2$ lattice points.
Saturating the path integral (the partition function) $Z$
by single-vortex configurations, we obtain\footnote{Our present discussion is merely 
qualitative and meant to highlight the 
competition between entropy and Boltzmann weights; exact treatments of the
Kosterlitz-Thouless transition are given in many textbooks, see, e.g.\ \cite{zinnjustin}.}
\beq\label{0.1}
Z \equiv \e^{-F/k_BT} =  
\!\!\!\!\sum_{{\rm spin~configurations}}\!\!\!\! \e^{-E[{\rm spin}]/(k_BT)}
\approx \left(\frac{R}{a}\right)^2 \; \e^{-[\k \ln(R/a)]/k_BT}.
\eeq
We note that the factor $(R/a)^2 $ is entirely entropic, simply arising from counting the
possible single-vortex configurations, and is independent
of any ``bare'' coupling constants (the spin coupling $\k$ and temperature 
$T$). Since the corresponding entropy 
$S= k_B \ln ({\rm number~of~configurations})$ has the same functional
form as the vortex energy, we can express the free energy as 
\beq\label{0.2}
F= E-ST = (\k -2k_B T) \ln (R/a). 
\eeq
The Kosterlitz-Thouless transition between a low-temperature 
phase (where vortices play no role) and a high-temperature phase 
(where vortices are important) occurs when $F=0$, i.e.\ when the entropy
factor is comparable to the Boltzmann weight of the classical energy.
At this point we are far from the na\"ive weak coupling limit of the 
lattice spin theory, which is just a Gaussian free field. Instead, the 
continuum field theory associated with the transition is the
sine-Gordon field theory at the coupling constant value where it 
changes from a super-renormalizable theory to a renormalizable 
one. 

The situation in four-dimensional CDT is analogous, in the sense that the 
nontrivial structure of the phase diagram 
reported in \cite{cdt-phasediagram} results from a competition between 
the entropy of configurations and the action, precisely as in \rf{0.2}. 
The analogy goes even further: 
when written as an effective action for the global scale factor (which plays
the role of an order parameter for the gravity case), and in the region of phase
space we have identified as possessing a meaningful classical limit, 
the free energy has the same functional 
form as the classical action, but with the opposite sign 
(corresponding to $F < 0$ in \rf{0.2}). This makes it important to understand
the semiclassical nature of this phase, and its relation to continuum physics. 
The investigation of this issue is the main
purpose of the present article.
  
The remainder of this article is organized as follows. In Sec.\ \ref{sec2} we
recall the set-up of causal dynamical triangulations and introduce a new
way of (approximately) fixing the total four-volume of spacetime in the simulations.
Using this new prescription, we re-analyze three-volume distributions in 
Sec.\ \ref{sec3}
and examine the probability distributions for different values of discrete 
three-volumes. The latter allows us to exhibit and 
quantify lattice artifacts for small
three-volumes, which occur close to the beginning and end of our
quantum universe. 
In Sec.\ \ref{sec4} we introduce a refinement of the spatial slicing, 
associated with connected layers of the different types of simplicial 
building blocks. This allows us to analyze the corresponding 
volume distributions separately, and compare their relative scaling
behaviour as a function of the bare coupling constants. In Sec.\ \ref{sec5}
we return to the task of reconstructing the effective action for the
three-volume fluctuations from the computer measurements. In a
nontrivial extension of
earlier work we look for evidence for the presence of corrections
to the effective action associated with curvature-squared terms in
the continuum. We summarize and discuss our findings in Sec.\ \ref{sec6}.
The appendix contains a extension of the analysis of Sec.\ \ref{sec5}
to the situation with a refined spatial slicing.

\section{Causal Dynamical Triangulations}\label{sec2}

In this section, we will review briefly some key ingredients of our approach 
and give an outline of the methods used. The basic motivation behind
quantizing gravity via CDT was explained in \cite{al}, and the current set-up
described earlier in \cite{ajld4,blp,bigs4}, to which we also refer for further
technical details.\footnote{An up-to-date reference describing
CDT and key results obtained in it is \cite{cdtlecturenotes}.}

Assuming that a path integral representation of quantum 
gravity exists, the basic idea is to provide it with an ultraviolet cut-off 
by using piecewise linear (``triangulated") geometries in the 
quantum superposition. Assigning lengths to the one-dimensional
edges of such a triangulation fixes its geometry completely, without
the need to introduce coordinates \cite{regge}, thus also avoiding
the redundancies of the usual continuum description of curved
geometries. For our purposes, we further restrict 
the piecewise linear manifolds to those which can be 
obtained by gluing together two specific types of building blocks 
(``four-simplices") with prescribed edge lengths. The typical edge
length serves as a geometric UV cut-off and is a measure of the 
fine-grainedness of the geometry.

Unlike its Euclidean counterpart, CDT also employs a proper-time foliation,
with respect to which the topology of space is not allowed to change. 
Admissible geometries are those which can be constructed
by first triangulating spatial slices of constant proper time,
which for simplicity we assume to have the topology of $S^3$. 
Each three-slice is assembled from identical
building blocks, namely, equilateral tetrahedra whose spacelike edges 
all have length $a_s$ (the {\it spatial} lattice spacing or UV cut-off). 
The gluing and the number of tetrahedra in each slice are arbitrary 
except for the overall topology and the imposition of local 
manifold constraints.\footnote{Note
that neither of these constraints on individual path integral histories will necessarily 
survive in the continuum limit. Previous results from CDT demonstrate that it can and
does happen that the system is driven dynamically to a quantum configuration
which is no longer of topology 
$S^3\times [0,1]$ and no longer resembles a four-dimensional manifold. 
This reflects the nontrivial interplay of ``energy" and ``entropy" already alluded to in the
previous section, which the nonperturbative CDT formulation allows us to capture.}
The next step consists in connecting neighbouring spatial slices
by timelike edges of length $a_t$ (the lattice spacing in {\it time} direction), 
in a way consistent with the ``filling" of the spacetime sandwiches by 
(up to rotations) two types of flat, Lorentzian four-simplices, to wit,
\begin{itemize}
 \item simplices of type (4,1), which have {\it four} vertices in common with the
spatial slice at (integer proper) time $n$ (thus spanning one of the equilateral
tetrahedra making up the slice) and {\it one} vertex with the slice at time $n + 1$. 
Time-reversed building blocks which share one vertex with slice $n$ and four
vertices with slice $n+1$ are ``of type (1,4)". 
\item simplices of type (3,2), which have {\it three} vertices in common with the
spatial slice at (integer proper) time $n$ (thus spanning one of the equilateral
triangles contained in the slice) and {\it two} neighbouring vertices with the slice 
at time $n + 1$, spanning a spacelike edge there.
Time-reversed building blocks which share two vertices with slice $n$ and three
vertices with slice $n+1$ are ``of type (2,3)". 
\end{itemize}
These four-simplices are glued pairwise along their three-dimensional ``faces", 
forming a layered, four-dimensional simplicial manifold of topology $S^3 \times [0,1]$
in the manner just described.   
In such a piecewise linear geometry the curvature is concentrated at the 
two-dimensional subsimplices (the triangles) and induces nontrivial
rotation angles on vectors which are parallel-transported around them.

Two neighbouring spatial slices labeled by integers $n$ and $n+1$
are separated by a proper-time distance $a_t$ in the sense that 
each timelike edge connecting the two hypersurfaces has this length.
Instead of using two lattice spacings $a_s$ and $a_t$, we usually
work with a single $a:=a_s>0$ and the dimensionless ratio 
$\alpha :=-a_t^2/a_s^2$. Expressing $\alpha$ in terms of squared
lengths allows us (i) to start out in Lorentzian signature,
where $a_t^2<0$, and thus $\alpha >0$, and (ii) to perform for each triangulation
an analytic continuation of $\a$ in the lower-half 
complex $\alpha$-plane to real, negative $\a$, resulting in 
a piecewise linear geometry of {\it Euclidean} signature and length 
assignments $a_t^2= |\a| a_s^2$ to the edges that in Lorentzian signature used
to be timelike. 
This ``Wick rotation" relies on the foliation in proper time. It has no obvious 
correspondence in the metric continuum formulation (see \cite{dl} for a discussion). 
Under this map, the Einstein-Hilbert action of a given Lorentzian
piecewise linear geometry, the so-called Regge action $S_L$ \cite{regge}, 
changes according to $S_L(\a)\to i S_{E}(-\a)$, where $S_E(-\a)$ is the 
Regge action for the corresponding piecewise linear Euclidean 
geometry with the length assignments $a_t^2= |\a| a_s^2$. 
A necessary condition to ensure that the Euclidean four-simplices
are nondegenerate is $|\a| >7/12$, which will be assumed in what follows. 
Once the rotation to Euclidean signature has been performed, we
redefine $\a \to -\a$, $S_E(-\a) \to S_E(\a)$ for simplicity of notation.

In our construction of the path integral, we start out with the set 
of causal piecewise linear Lorentzian geometries described above.
In order to perform the sum over these histories, we then rotate each of them
to Euclidean signature, so that it will contribute with a real weight 
$\exp (-S_E)$. More precisely, we have 
\bea
&&S_E^{cont}= \frac{1}{16\pi G} \int \sqrt{g} (-R+2\La) \nonumber \\
 &\to & S_E=-(\kappa_0+6\Delta) N_0+\kappa_4 (N_{4}^{(4,1)}+N_{4}^{(3,2)})+
\Delta (2 N_{4}^{(4,1)}+N_{4}^{(3,2)}),
\label{actshort}
\eea 
where $N_0$, $N_4^{(4,1)}$ and $N_{4}^{(3,2)}$ are the total numbers 
of vertices (zero-simplices), of four-simplices of {\it both} type (4,1) and type (1,4),
and of four-simplices of types (3,2) and (2,3) respectively. 
The Regge action \cite{regge}
takes the particularly simple form exhibited in the bottom line of eq.\ \rf{actshort}
because each CDT contains only two geometrically distinct 
types of building blocks.
The parameter $\k_0$ in (\ref{actshort}) is proportional to the inverse bare
gravitational coupling constant, $\k_4$ is a linear function of the bare
cosmological and the inverse bare gravitational constant, while
$\Del$ is an asymmetry parameter, which for given $\k_4$
and $\k_0$ can be related to $\a$. It is normalized such that 
$\a=1$ (the case of equilateral Euclidean four-simplices)
corresponds to $\Del=0$. 

For technical reasons we perform the numerical simulations at fixed
four-volume, which in practice is usually realized by fixing
$N_4^{(4,1)}$ to some target volume $V_4$. 
The geometry is updated by using local Monte Carlo moves 
(c.f. \cite{ajld4,blp,bigs4}), 
which in general will change the four-volume.
In past papers this was taken care of by modifying 
the action by a linear term according to
$S_E \to S_E + \eps |(N_4^{(4,1)}-V_4)|$, with a parameter $\eps$
controlling the range of fluctuations, and
measurements collected only when $N_4^{(4,1)} = V_4$ exactly. 
Fixing $N_4^{(4,1)}$ in this delta function-like manner has the side
effect of generating a zero mode in the correlation matrix 
$\langle \delta V_3(n) \delta V_3(n')\rangle$ of fluctuations
in the spatial {\it three}-volume $V_3$ in time, an important quantity in investigating
the dynamical behaviour of the quantum universe \cite{bigs4}.
Projecting out the zero mode, which is necessary for inverting this
matrix, leads to an inconvenient mixing of the remaining modes. 
To get around this in the present paper, we no longer 
require that $N_4^{(4,1)} = V_4$ exactly in measurements,
but instead let it fluctuate according to the modified action
$S_E + \eps (N_4^{(4,1)}-V_4)^2$, which eliminates the zero mode. 
The advantage of using a quadratic term is that it can be treated easily 
alongside other terms in the action.

In a given simulation, the parameter $\kappa_4$ of the model is fixed by
requiring that $\langle N_4^{(4,1)}\rangle  = V_4$. 
The data presented below were taken at $\eps=10^{-5}$, but we checked that 
the results are essentially unaltered for  
$\eps=2\times 10^{-5}$ and $\eps=5\times 10^{-6}$. 
Furthermore, we also checked (at $\eps=2\times 10^{-5}$)
that using the total number $N_4$ of four-simplices instead of $N_4^{(4,1)}$ 
as the target volume does not change the situation either.
Ideally one would like to make $\eps$ as small as possible, but this
must be balanced against the fact that 
smaller $\epsilon$-values increase the auto-correlation time.

Depending on the values of the bare couplings in (\ref{actshort}), the 
CDT theory will appear in one out of three different phases A, B or C
(see Fig.\ \ref{Fig5a} for a depiction of the phase diagram in the
$\k_0$-$\Delta$-plane, 
and \cite{cdt-phasediagram} for a detailed description). 
In the present paper we will concentrate on phase C
where an extended de Sitter universe has been 
observed \cite{agjl,bigs4,cdt-phasediagram}. 
In this phase we can construct an effective semiclassical action 
for the scale factor of the universe. Changing  
the bare coupling constants $\k_0$ and $\Del$ will affect
the effective coupling constants appearing in this action.

\section{Volume distributions reloaded}\label{sec3}

\begin{figure}
\centering
\includegraphics{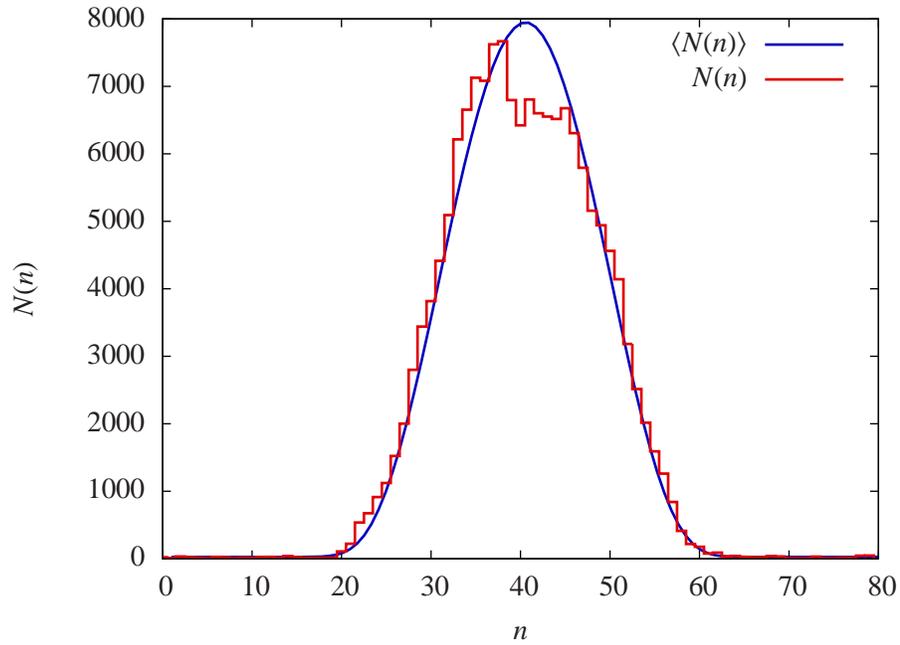}
\caption{Example of a volume distribution $N(n)$ for a specific
path-integral configuration (red), compared to the average 
distribution $\langle N(n)\rangle$ (blue). On the scale of the plot, the latter
cannot be distinguished from a fit to the theoretical curve of 
relation (\ref{sphere}).}
\label{Fig1}
\end{figure}

An interesting observable investigated previously 
is the typical ``shape" or ``volume profile" 
of the quantum universe, more precisely, the average distribution  
\bea
\langle N(n)\rangle := \langle N_4^{(4,1)}(n)\rangle
\eea
of spatial volume as function of the discrete proper 
time $n$.\footnote{Equivalently, one can work with the number 
$N_3(n)$ of three-simplices, where $N_4^{(4,1)}(n)=2 N_3(n)$
for all interior constant-time slices of the triangulation.}
What we found is that individual configurations 
in the well-behaved ``de Sitter phase" have a characteristic shape, 
consisting
of a (spatially extended) bulk contribution or ``blob" and a separate, thin ``stalk", 
whose spatial volume for all times stays close to the prescribed
minimal cut-off size.
Fig. \ref{Fig1} shows an example of the volume distribution 
$N(n)$ of a typical, individual path integral configuration, 
compared to the average distribution $\langle N(n)\rangle$.

If we compactify the time direction, as is usually done in simulations,
the action is symmetric under time translations,
permitting us to move the position of the ``centre of volume'' 
along the time direction.
To determine the average distribution we shift this position for every
individual configuration such that its centre coincides with a
common reference point $n_0$ on the time axis. For each shifted configuration
we measure the three-volume distribution $N(n)$ 
obtained in numerical simulations
at some fixed $V_4$ as a function of the shifted time variable $n$,
which {\it can} be averaged meaningfully
over many independent configurations.
In \cite{semi} it was shown that inside 
the blob region the average distribution obeys the characteristic 
scaling relation
\bea
\langle N(n)\rangle = A_s V_4^{1-1/d_H} \cos^{d_H-1}(\tau),
\label{sphere}
\eea
where $\tau = A_t n/ V_4^{1/d_H}$ represents proper time and $d_H = 4$ 
(within numerical accuracy), and $A_t$ and $A_s$ are numerical constants. 
In the stalk part the average volume
is practically constant and independent of $V_4$. 
The fact that the numerical data reproduce the distribution (\ref{sphere}) 
with great accuracy remains 
practically unchanged when the measurements 
are performed with the new method.

The interpretation of the above results is that the quantum system 
generates the three-volume distribution of a round four-sphere (i.e.
Euclidean de Sitter space) with
four-volume $V_4$, with time steps proportional to the 
(cosmological) proper time separating the spatial hypersurfaces.
The stalk is present only because (i) we have fixed the time 
period $T$ to be larger
than the proper-time extension of the universes created, and
(ii) we enforce a minimal three-volume of 5 tetrahedra per spatial slice, 
corresponding 
to $N_{\rm min}(n)=10$, in keeping with the simplicial {\it manifold} character
of the spacetime.
Although it appears from the simulations that the dynamics of the system 
wants to drive the three-volume to zero 
at the two ends of the blob, we prefer to maintain these kinematical
restrictions (and thus the stalk region), because they allow us to monitor 
fluctuations in the time extent
of the blob, which would be suppressed if we tried to adjust the
time interval $T$ to match the blob exactly.  

The measurements presented below were taken for the coupling constant
values $(\kappa_0,\Delta)=(2.2,0.6)$, 
four-volume $ V_4 = 160~000$ and time period $T=80$, and exhibit a 
behaviour typical for systems inside the de Sitter phase. 
The reference time has been fixed to 
$n_0=40.5$ and the distribution symmetrized
with respect to $n \to 81-n$. Comparing the volume distributions  
with relation (\ref{sphere}), the range of the blob is roughly $17 < n < 64$ in
terms of integer times, whereas the stalk region is located in 
$1 \leq n \leq 17$ and $64 \leq n \leq 80$. 

\begin{figure}
\centering
\includegraphics[width=0.85\textwidth]{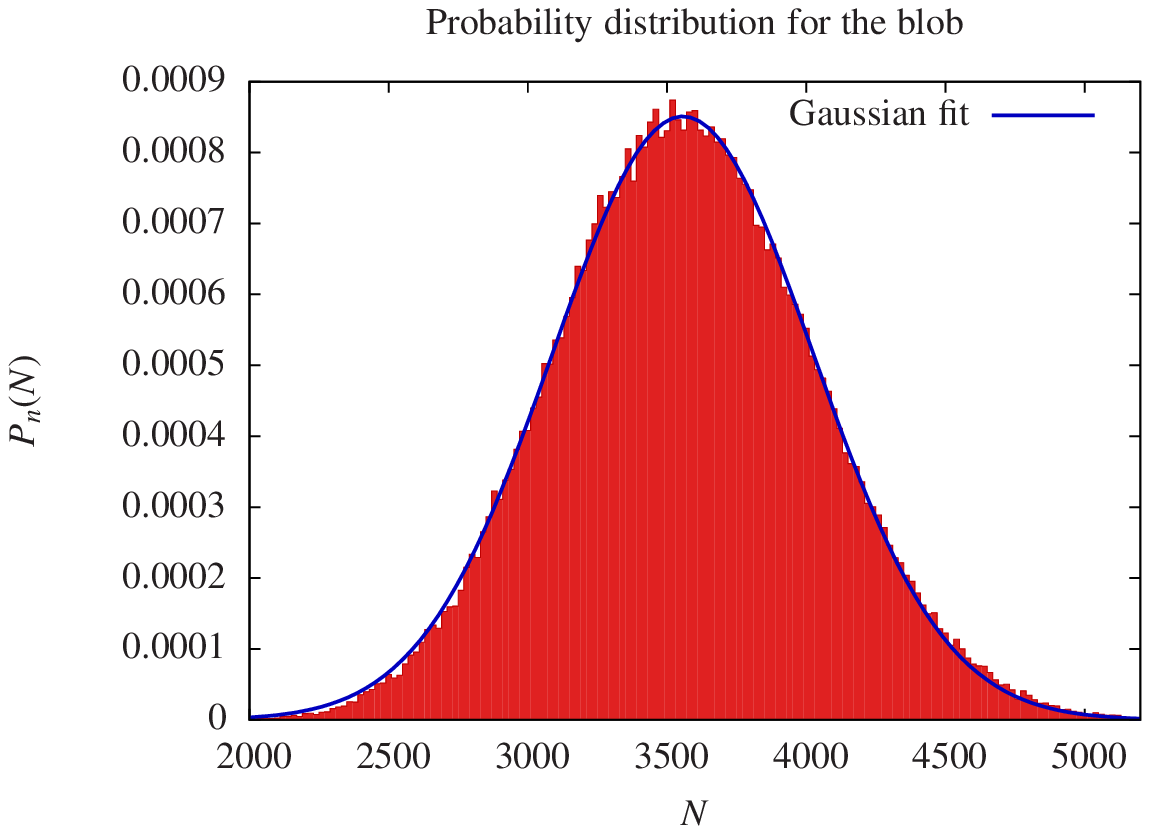}
\includegraphics[width=0.85\textwidth]{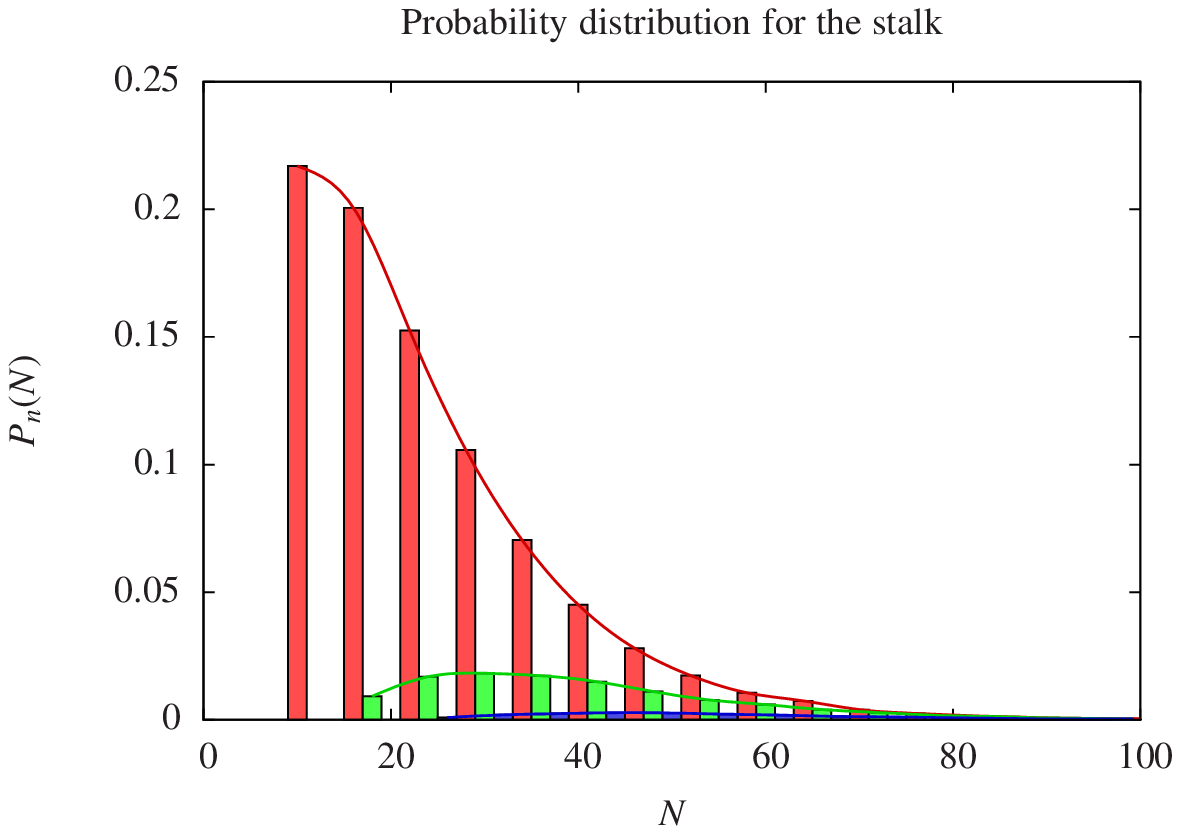}\\
\caption{Top: probability distribution ${\cal P}_n(N)$ of three-volumes 
at fixed time $n=29$; 
bottom: probability distribution ${\cal P}_n(N)$ in the stalk (for every $n\leq 17$);
the data fall into three families (colour-coded in the graph), 
each one with a different behaviour.}
\label{Fig2}
\end{figure}

In previous work \cite{semi,agjl,bigs4} we have matched the average 
$\langle N(n)\rangle$ to the classical solution 
$N_{cl}(n)$ of a (discretized) mini-superspace action. 
This semiclassical interpretation is corroborated further by 
new measurements made
of the probability distributions 
${\cal P}_n(N)$, which for a given time $n$ describe the probability that the 
three-volume $N(n)$ is equal to $N$. 
If the semiclassical picture is indeed correct, 
these distributions should be approximately
Gaussian with a mean equal to $\langle N(n)\rangle$ and a dispersion 
$\sigma_n^2 = \langle N(n)^2\rangle-\langle N(n)\rangle^2$.
This is exactly what we observe in the range $25 \leq n \leq 56$, 
well inside the blob,  
as shown in Fig.\ \ref{Fig2} (top).

Inside the stalk, the situation is completely different.
Not only the average $\langle N(n)\rangle$, but also the three-volume distribution
${\cal P}_n(N)$ is unchanged when we move along the stalk. 
In addition, we observe short-distance lattice artifacts,
in the form of a split of the distribution of three-volumes 
into three separate families. 
This is clearly visible in Fig.\ \ref{Fig2} (bottom), which
shows the average three-volume 
distribution in the stalk for $1\leq n \leq 17$.
Inside each family the discrete three-volumes differ by 6, such that we have three
sets $\{10, 16, 22, \dots\}$,
$\{12,18,24,\dots\}$ and $\{14,20,26,\dots\}$.\footnote{Note that 
by definition $N(n)$ is always an even number.}  
Although the presence of the stalk -- as we have argued above -- is 
merely a consequence of our kinematical set-up, it cannot simply be
ignored, but has to be taken into account in the calculations, for example,
when analyzing the covariance matrix of three-volume 
fluctuations and its inverse. In order to distinguish short-distance artifacts
from short-distance physics, it is therefore important to understand the
behaviour of the transition region between stalk and blob as best possible.

\begin{figure}
\centering
\includegraphics[width=0.85\textwidth]{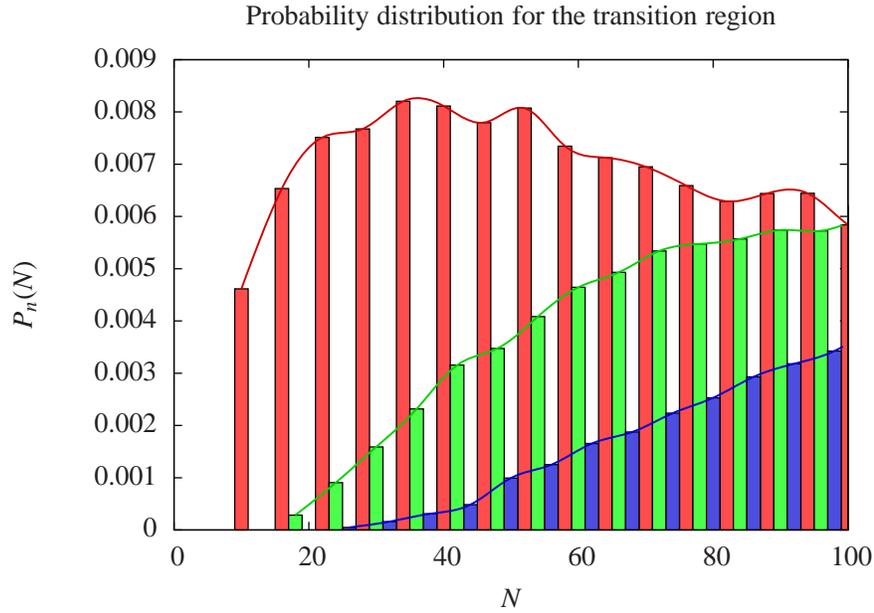}\\
\includegraphics[width=0.85\textwidth]{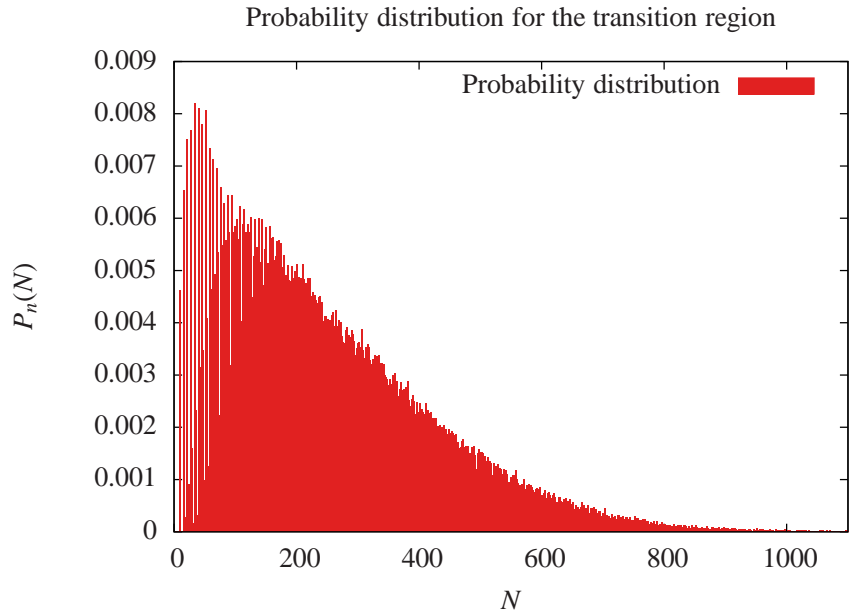}
\caption{The probability distribution $P_{22}(N)$ from the transition 
region near the end of the blob. 
For small $N$ the distribution splits into 3 families (top). 
For large $N$ the split disappears, but the distribution 
is highly asymmetric (bottom, no colour-coding for the three families). }
\label{Fig3}
\end{figure}

An analogous split into three sets of the three-volume distribution 
for small volumes is also observed in the region $17<n<25$,
which smoothly joins the stalk and which we have classified as 
``inside the blob", but where $\langle N(n)\rangle$ is still relatively small.
We find that
for $10 \leq  N(n)  < 200$ the distribution again splits into three families. 
An example for $n=22$ is shown in Fig.\  \ref{Fig3} (top); the situation is 
similar for $18 \leq n \leq 24$.
If in the same transition region one considers higher values of 
$N(n) \gtrsim 200$, the split between the three families disappears, 
but one also notes that the distributions are highly asymmetric 
and non-Gaussian, as again illustrated for $n=22$, depicted
in Fig.\ \ref{Fig3} (bottom).

We conclude that lattice artifacts are visible for spatial slices of sizes
up to $N_3=100$ tetrahedra (corresponding to $N=200$). This may
sound like much, but in reality corresponds to rather small linear
distances. This can be understood by comparing the total volume of
100 spatial tetrahedra, $100\times \frac{\sqrt{2}}{12} \, a^3_s$, with
that of a regular three-sphere, $2 \pi^2 R^3$. If we arranged the
tetrahedra to approximate such a sphere, we would have $R \approx a_s$,
i.e. only a single lattice spacing, with the antipodal 
distance on the three-sphere being roughly equal to $3a_s$. 
Viewed like this, it appears rather encouraging that the observed 
lattice artifacts vanish so quickly as function of the three-volume. 

Quantifying the short-distance artifacts for one quantity
gives us a good idea of the scale at which they occur for a
given four-volume, but still leaves us with the difficulty of disentangling
them from genuine quantum-gravity signatures in this and other
observables.
For example, the non-Gaussian character of the distribution of
fluctuations around the sphere \rf{sphere} at small $N_3(n)$
observed above could indicate new short-distance physics, say,
along the lines suggested in the asymptotic safety scenario
\cite{reuter}. We presently do not know how to relate the
deviation from the four-sphere at small scale factor $a(\tau)$
described there to our explicit small-$N(n)$ ``observations'', but it would 
clearly be interesting to do so.

\section{Refining the spatial slicing}\label{sec4}
\begin{figure}
\centering
\includegraphics[width=0.85\textwidth]{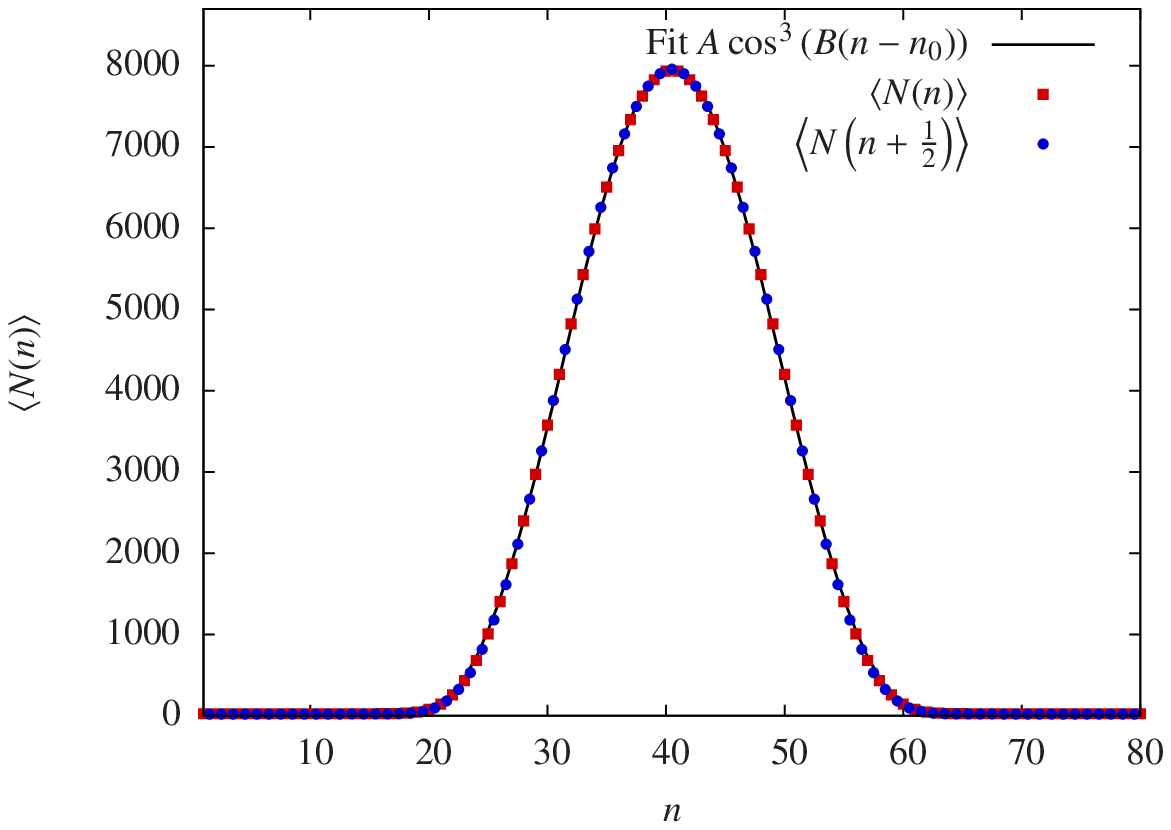}
\caption{The averaged distributions $\left\langle N(n) \right\rangle$ and 
$\left\langle N\left(n + \frac{1}{2}\right) \right\rangle$ combine into a 
single curve after performing a suitable relative rescaling. 
}
\label{Fig4a}
\end{figure}

An alternative description of the four-dimensional simplicial manifolds
contributing to the sum over histories is given in terms of their dual structures,
where we assign vertices to the centres of four-simplices and links to the 
three-dimensional faces between adjacent four-simplices. 
In the absence of boundaries of the manifold, each dual vertex will
have five neighbouring ones, connected by dual links. This dual picture
suggests a further subdivision of the original (integer) time steps of the discrete
triangulation, as we will now explain. 

Integer steps $n$ provide natural units for timelike paths in the
original lattice if the paths are taken to run only along the timelike\footnote{``timelike" 
refers to the character of the 
geometric elements {\it before} the Wick rotation} edges of
the four-simplices, which form
one-dimensional `connectors' between pairs of
adjacent constant-time layers made entirely from spatial
tetrahedra. 
Constructing analogous paths on the dual lattice,
at least four steps are required to 
connect a vertex dual to a (4,1)-simplex at time $n$ with a vertex dual to a (4,1)-simplex 
at the next time step $n+1$. Labelling the dual vertices by their associated simplex types, such
a path is given by a sequence $(4,1) \to (3,2) \to (2,3) \to (1,4) \to (4,1)$. 
In addition, we have observed that all dual
vertices of types (3,2) and (2,3) between times $n$ and $n+1$ 
and the dual links connecting them 
form a single closed, connected graph.
This makes it natural to assign a time $n+1/2$ to this
layer and study the properties of the distribution of
\bea
 \bar{N}(n+1/2) := N_4^{(3,2)}(n+1/2),
\eea
where $N_4^{(3,2)}(n+1/2)$ is the total number of (3,2)- and (2,3)-simplices located 
between discrete times $n$ and $n+1$. It is not surprising that this distribution is  
similar to $\langle N(n)\rangle$, in fact, by a simple rescaling
\bea
N(n+1/2) = \rho \bar{N}(n+1/2)
\eea
(where the constant $\rho$ will depend on the values of the couplings),
one can achieve that in the blob range the combined distribution $\langle N(n)\rangle$, with $n$ now
running over both integer and half-integer times, 
is well approximated by a single,
smooth curve according to formula (\ref{sphere}). 
Note that this unification is not global, in the sense that a different rescaling
is needed in the stalk part, which is not surprising in view of the different dynamics in
this region. Fig.\ \ref{Fig4a} shows the resulting distribution 
for the standard choice of bare couplings, $\k_0=2.2,~\Del=0.6$. 

\begin{figure}
\centering
\includegraphics[width=0.85\textwidth]{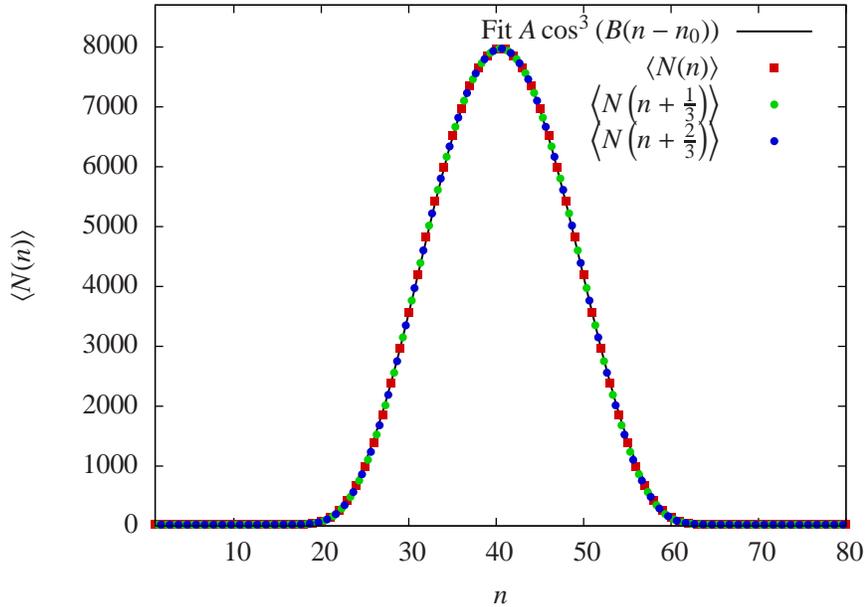}
\caption{Combining the averaged volume distributions $\left\langle N\left(n\right) \right\rangle$, 
$\left\langle N\left(n + \frac{1}{3}\right) \right\rangle$ and $\left\langle N\left(n + \frac{2}{3}\right) \right\rangle$
into a single one.}
\label{Fig4b}
\end{figure}
One can push this line of argument further by performing an even finer discrete
subdivision of time.
As argued above, going forward in time along dual links necessarily takes one through
a sequence of different types of four-simplices. Nothing new can be learned by distinguishing
between a layer of (4,1)-simplices and that of the adjacent (1,4)-simplices, because they
are in one-to-one correspondence (each interior spatial tetrahedron at integer $n$ is shared
by exactly one (4,1)- and one (1,4)-simplex). 
By contrast, 
there is no such relation between the (3,2)- and the (2,3)-simplices.
This has motivated us to divide each integer time step into three and
study volume distributions separately for the sets of (4,1)-, (3,2)- and
(2,3)-simplices contained in each `sandwich' $[n, n+1]$.
One finds again that they can be combined into a single, universal distribution 
by defining
\bea
 N(n+1/3) := 
2\rho N^{(3,2)}(n+1/3),\quad  N(n+2/3) :=
2\rho  N^{(2,3)}(n+2/3), 
\label{drittel}
\eea 
with the same $\rho$ as used before. In (\ref{drittel}),
$N^{(3,2)}(n+1/3)$ and $N^{(2,3)}(n+2/3)$ count
the number of (3,2)- and (2,3)-simplices, which collectively have
been assigned the time labels $n+1/3$ and $n+2/3$, respectively. 
Fig.\ \ref{Fig4b} illustrates that the sphere fit continues to work beautifully
also with respect to this time subdivision. These results underscore that
the detailed choices we make at the level of the individual building blocks,
in the present example the identification of microscopic time steps, bear little
direct relation to the macroscopic aspects of the semiclassical emergent
geometry. This may have been anticipated because of the 
nonperturbative nature of this limit in CDT quantum gravity. 

\begin{figure}[t]
\centering
\includegraphics[width=0.75\textwidth]{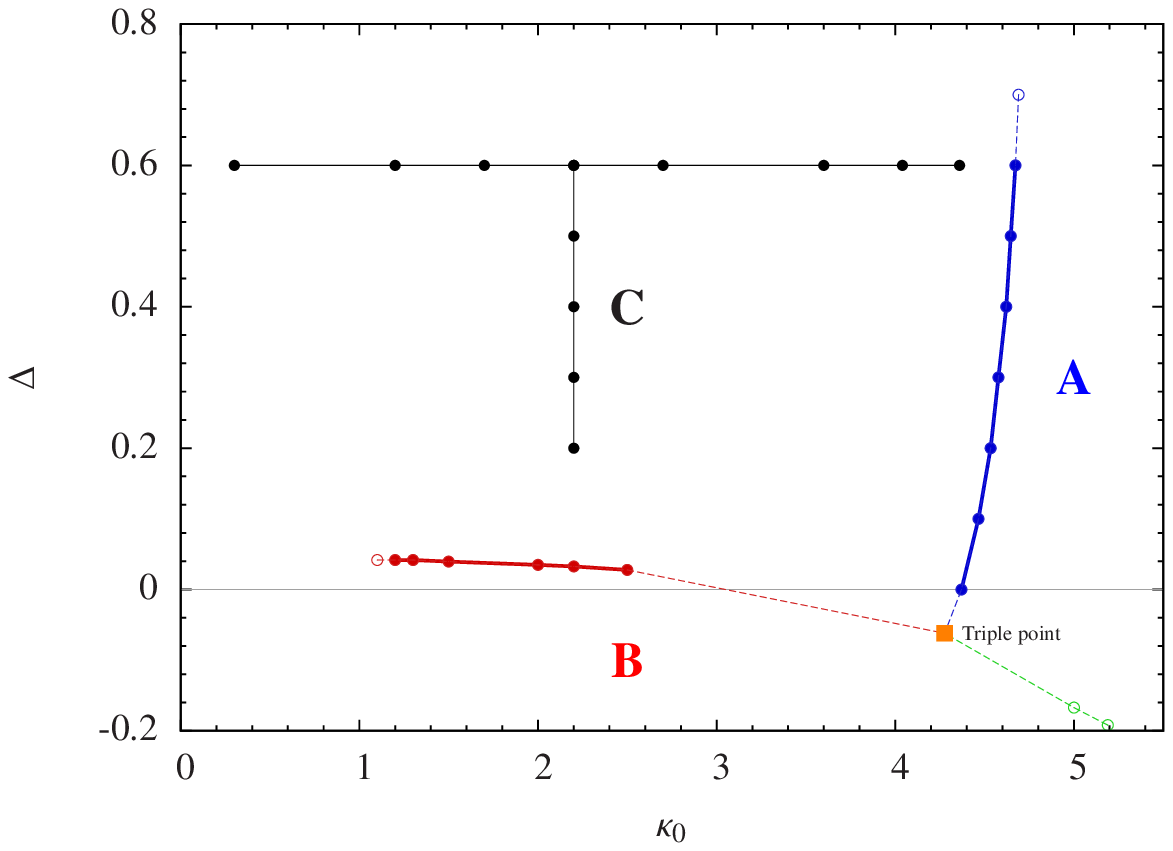}\\[1ex]
\includegraphics[width=0.45\textwidth]{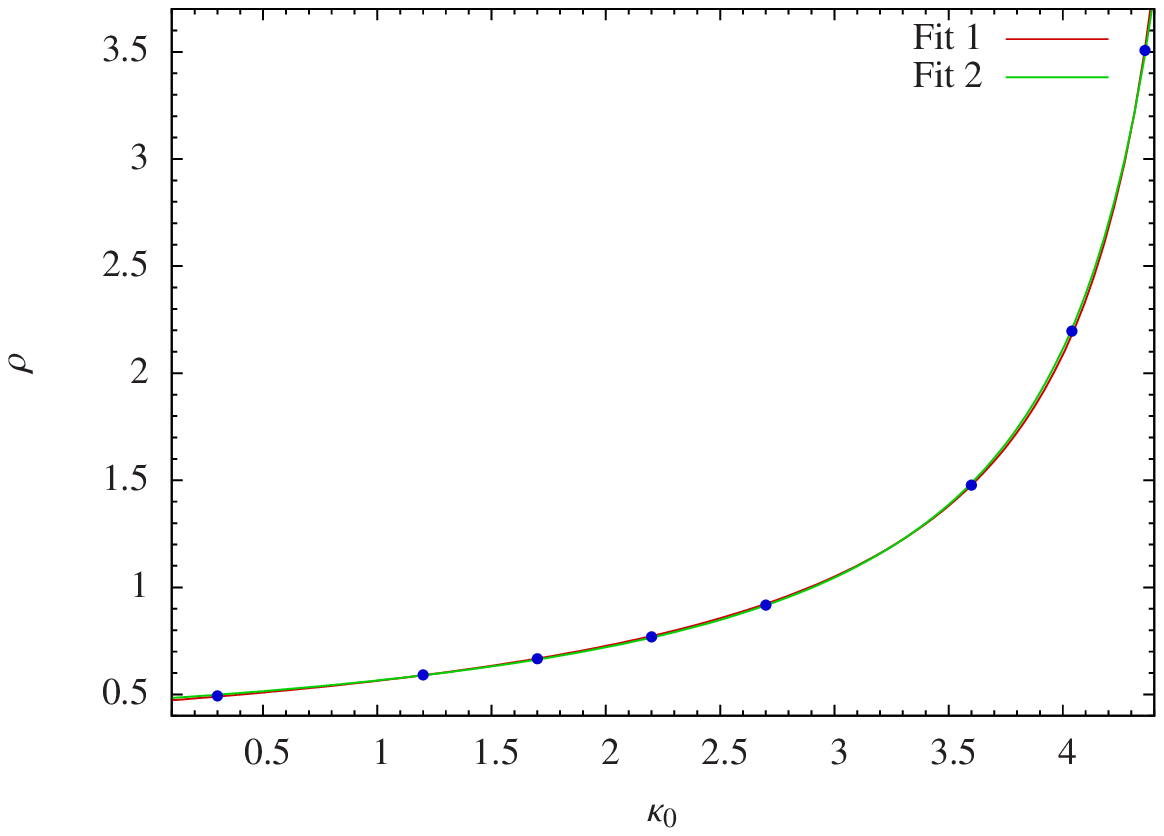}
\includegraphics[width=0.45\textwidth]{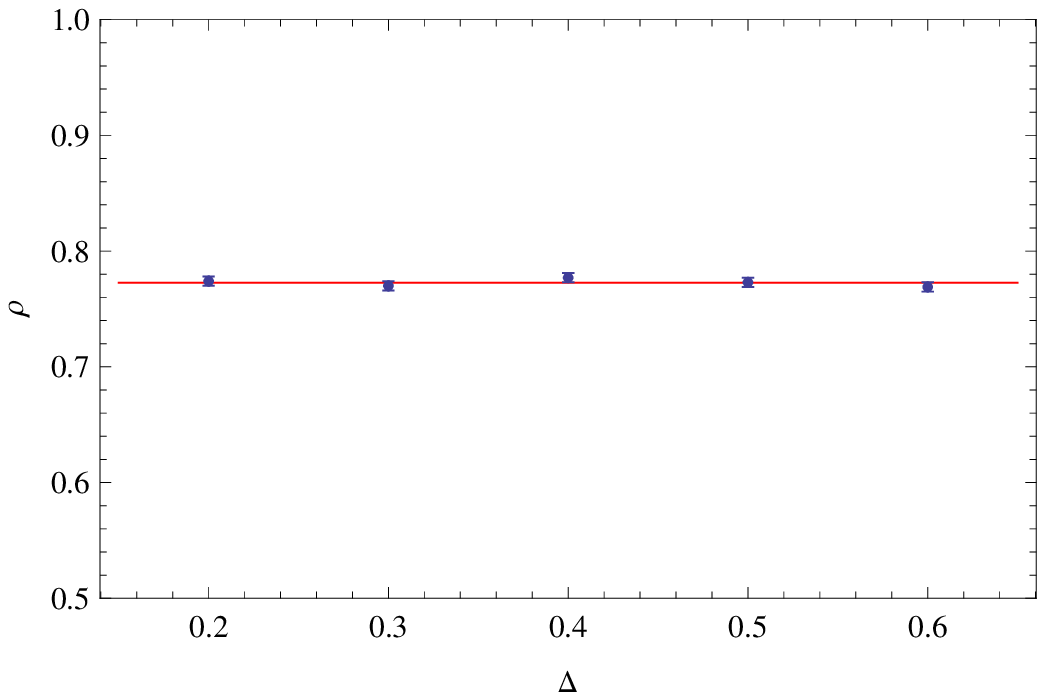}
\caption{Systematic study of the dependence of the parameter
$\rho$ on the couplings $\k_0$ (bottom left) and
$\Delta$ (bottom right); the points in the phase diagram 
at which the measurements were made lie along the T-shaped
line in phase $C$ (top figure).}
\label{Fig5a}
\end{figure}

Our next step will be to repeat the above measurement for different
values $(\k_0,\Del)$ of the bare coupling constants, and to study systematically
how the scaling parameter $\rho$ behaves as a function $\rho=\rho(\k_0,\Del)$.
We are particularly interested in the behaviour inside the de Sitter phase (labelled
C in Fig.\ \ref{Fig5a}, top) as we approach one of the critical lines, which separate
phase C from phases A and B. 
The set of points in the coupling constant-plane where measurements were taken 
lies along the T-shape inside phase C.

As is apparent from Fig.\ \ref{Fig5a}, bottom right, 
there is practically no dependence of $\rho$ on $\Delta$ for fixed $\k_0$,
despite the fact that the quality of the fits to the functional form (\ref{sphere})
deteriorates for $\Delta < 0.3$. On the other hand, leaving $\Delta$ fixed
and varying $\k_0$, the dependence becomes quite pronounced, with
$\rho$ increasing monotonically as the A-C phase transition is approached. 
However,
with the current data it is not really possible to nail down uniquely
the functional form of the dependence, and thus the nature of the phase
transition. Let us illustrate this point by drawing two different
three-parameter fits through the data 
points of Fig.\ \ref{Fig5a}, bottom left.  
(Note that the error bars of the data points themselves are too small to be visible in
the figure.) For fit 1 (red curve) we choose the form
\beq\label{fit1}
\rho = A (\k_0^c -\k_0)^{-C},
\eeq
for which we have determined the free parameters as 
\beq\label{fit1a}
A        = 1.67 \pm 0.03,~~
\k_0^{c} = 4.76 \pm 0.02,~~
C        =0.82\pm 0.03~~~(\chi^2 \approx 1). 
\eeq
Given the excellent agreement with the data and the small error bars, one
is tempted to conclude that the transition is characterized by a
fractional critical exponent, a result which 
usually indicates a higher-order phase transition.
However, such a conclusion would be premature. The error bars
of such fits can be misleading, since they refer to a specific functional form, 
which is unlikely to be correct when we move away from the critical line.
As an alternative, let us assume that $1/\rho$ is analytic in $\k_0$
and goes to zero linearly as 
we approach the critical value $\k_0^c$, instead of with a fractional
power. 
Matching the number of free parameters, the corresponding
fit 2 (green curve) has the form
\beq\label{fit2}
1/\rho(\k_0) = A + B * \k_0 + C * \k_0^2, 
\eeq
and its parameters have been determined as
\beq\label{fit2a}
A =  2.10 \pm 0.02,~~
B = -0.30 \pm 0.02,~~ 
C = -0.026 \pm 0.003~~~(\chi^2 \approx 2), 
\eeq
corresponding to a critical value $\k_0^c = 4.86\pm 0.22$.
We note that this fit is almost as good as the previous one 
(with $\chi^2 \approx 2$ rather than $\chi^2 \approx 1$), 
and the two curves are barely distinguishable on the plot. However, it leads
to the rather different conclusion that the
critical exponent is $-1$ rather than $-0.82$, indicative of a 
first-order transition. The fact that the exponent $-1$ lies more than 
5 standard deviations away from $-0.82$ reiterates our earlier
assertion that the error bars given with the individual fits 
cannot be taken seriously as the only source of error.
Deciding between these two possibilities will require taking data closer to the 
critical line, which is quite difficult from a simulation-technical point of view.

To summarize, the data presented above do not allow us to distinguish 
between a first- and a higher-order transition. 
The possibility of a higher-order A-C transition is intriguing,
since it is exactly by changing $\k_0$ that we encounter a
transition scenario \`a la Kosterlitz-Thouless as described 
in the introduction, however, it is not supported by the  
results reported in \cite{cdt-phasediagram}. There we observed 
an abrupt transition and clear evidence of a hysteresis 
when changing the coupling constants, which we interpreted
as good evidence of a first-order transition.
The results reported above do not really contradict this hypothesis,
although they leave the door ajar to the 
possibility of a higher-order transition.
More work is needed to settle this question decisively.

\section{Effective action: curvature corrections}\label{sec5}

As mentioned above, the measured averaged volume distribution 
$\la N(n)\ra $ closely matches the behaviour of the same quantity derived
from a discretized mini-superspace action.
In \cite{bigs4}, we made an effort to push this line of argument
further by trying to determine the form of the effective action from  
measurements of the covariance matrix of the three-volume fluctuations 
$\delta N(n) = N(n)-\langle N(n)\rangle$.  
Extracting these data was more subtle, in part due to the presence
of a zero-mode, whose origin was the constraint of keeping the number of 
(4,1)-simplices constant.
This mode had to be projected out before one could invert the fluctuation 
matrix and use it to reconstruct an action. An unpleasant feature 
of this projection is its mixing of effects 
from the stalk with ``bulk physics''. One new ingredient in the current work
is to lift the constraint of a constant number of (4,1)-simplices, and
thus avoid the zero-mode problem.

In this new setting, we study the covariance matrix 
$C_{n,n'}= \la \delta N(n) \del N(n')\ra $
and invert it to a matrix $S_{n,n'}$ which in principle describes
the (nonlocal) effective action resulting from fluctuations of the
three-volume around the 
semiclassical solution $N_{cl}(n)\equiv \la N(n)\ra $ (see \cite{agjl}
for details).  Because  we have added a term 
$\eps\left(\sum_n N(n) - V_4\right)^2$ to the original Einstein-Hilbert 
action, each matrix element of $S_{n,n'}$ receives a 
contribution $2 \eps$. 
After subtracting this shift, we find that 
all matrix elements outside the diagonal and the
neighbouring sub- and superdiagonals are zero up to numerical noise.
To quadratic approximation in the fluctuations we have 
\beq\label{derivative} 
S_{n,n'} =
\frac{\partial^2 S_\textit{eff}}{\partial N(n)\partial N(n')}\Big|_{N(n)=N_{cl}(n)}.
\eeq
Our observations then suggest 
that the effective action is quasi-local in time and can be expressed as
\bea
S_\textit{eff} = \frac{1}{\Gamma} \sum_n \Big( f\big(N(n),N(n+1)\big) + 
V(N(n)) \Big)+\eps\big(\sum_n N(n) - V_4\big)^2, 
\label{seff}
\eea
where the function $f$ and the potential $V$ need to be determined from the data
and $\Gamma$ is an overall constant.
The simplest function compatible with the observed scaling is
\bea\label{firstterm}
f(x,y) = \frac{(x-y)^2}{x+y}.
\eea 
Assume now that the three-volume (equivalently, the third power
$a^3(n)$  of the scale factor of the universe) behaves according to
\beq\label{functional}
\la N(n) \ra = A_s  V_4^{3/4} H\Big(\frac{n-n_0}{ V_4^{1/4}}\Big)
\eeq
for some function $H$,
where $n_0$ denotes the location of an origin chosen along the discrete time axis.
Comparing with \rf{sphere}, this scaling is consistent with our data
and we have
\bea\label{newsphere}
N_{cl}(n) =  A_s  V_4 ^{3/4} \cos^3(A_t (n-n_0)/  V_4^{1/4}),~~~~
A_s = \frac{3}{4}A_t,
\eea
where the last equality is required by normalization. 
Converting the function $f(N(n+1),N(n))$ found in \rf{firstterm} to
a continuum expression yields
\beq\label{firstterm1}
f(N(n+1),N(n)) \to
\frac{\rm const.}{N(t)} \left(\frac{d N(t)}{d t}\right)^2,
\eeq
where $t \propto n \, a_t/ V_4^{1/4}$ is a dimensionful time variable, 
and $a_t$ the lattice spacing in time direction.

We recognize (\ref{firstterm1}) as the kinetic term of a  
minisuperspace action for a spatially homogeneous and isotropic universe. 
One could consider a variety of corrections to this term. More
specifically, we are looking for corrections related to the 
short-distance behaviour of the theory. 
Taking guidance from the continuum theory of a homogeneous, isotropic universe
with (Euclideanized) four-metric
\begin{equation}
\label{unimet}
ds^2=dt^2+a(t)^2d\Omega_{(3)}^2,\;\;\; d\Omega_{(3)}^2=d\theta^2 +\sin^2\theta
(d\phi_1^2+\sin^2\phi_1\, d\phi_2^2),
\end{equation}
the most general minisuperspace 
action containing both the Ricci scalar and a curvature-squared term\footnote{For
a classical universe with metric (\ref{unimet}), the a priori distinct terms $R^2$,
$R_{\mu\nu}R^{\mu\nu}$ and $R_{\mu\nu\rho\sigma}R^{\mu\nu\rho\sigma}$ 
in the action are 
all proportional.} (the latter
with coupling constant $\omega$), as well as a cosmological-constant term is of the
form
\begin{equation}
\label{r2action}
12 \pi^2 \int dt\left[ \frac{1}{G}(-a\dot a^2-a+\frac{\Lambda}{3} a^3)
+\omega \left(\frac{1}{a}-2\frac{\dot a^2}{a}+\frac{
\dot a^4}{a}+a\ddot a^2\right) \right]. 
\end{equation}
We note the presence of fourth-order time derivatives, whereas the higher-order
spatial derivatives in such a universe are
converted to inverse powers of the scale factor $a(t)$. The $R$-term multiplied by
the inverse Newton constant in the integrand of (\ref{r2action}) contains also
a potential part $ a\propto N^{1/3}$, which in earlier work \cite{bigs4}
was matched successfully to computer measurements. The fact that 
corresponding terms in the effective action (\ref{seff}) appear with the opposite
sign compared to (\ref{unimet}) has to do with nonperturbative contributions
arising from integrating over the remaining degrees of freedom in the path
integral \cite{blp,semi,bigs4,cdtlecturenotes}, and is therefore another
feature of the nonperturbative nature of CDT's semiclassical limit in
phase C,  

For the purposes of our present investigation, do we observe any trace of the
terms contributing to the $R^2$-term in our effective 
action? 
With regard to the purely diagonal potential term $V(N(n))$ in (\ref{seff}), we would expect 
higher-order corrections\footnote{Such a power expansion should be valid as long
as $N(n)$ does not become small, i.e. sufficiently far away from the 
``beginning" and ``end" of the universe.} in powers of $N^{-2/3}$, that is,
\bea
V(N(n)) = - \lambda_\textit{eff} N(n) + \mu N(n)^{1/3}
+\xi_1 N(n)^{-1/3}+ {\cal  O}(N(n)^{-1})
\label{poteff}
\eea
to the order we are considering, where the first term is a Lagrange multiplier 
term.\footnote{From expression \rf{poteff} it is clear that if the first two
terms should both contribute to leading order in the continuum limit, at least
one of the multiplying constants needs to scale nontrivially as a function of $N$.
This is borne out by relation \rf{volexpand} below.}
Next, let us turn to the kinetic term in $S_\textit{eff}$.
Since we have already seen that only
nearest-neighbour terms contribute appreciably to the measured
covariance matrix, 
we can immediately conclude that a term corresponding to 
$a(t)\ddot a(t)^2$ -- whose natural discretization would contain a term 
like $(N(n+1)-2 N(n)+N(n-1))^2/N(n)$, and therefore 
next-to-nearest-neighbour interactions -- is absent. This is our first indication
that in the region of phase space under investigation, there is no appreciable
contribution from a squared-curvature term matching the one in the
continuum expression (\ref{r2action}), when re-expressed as
function of the three-volume $N$.

Let us look at another kinetic term expected to contribute to the
action at this order, $\dot a^4/a\propto \dot N^4/N^3$, 
to illustrate how a more quantitative analysis needs
to proceed and what the potential implications are for the continuum limit of the
theory, if we manage to construct one. 
Pretending for the time being that this term is the only correction to
the finite-difference expression (\ref{firstterm}) at this order leads to a modified ansatz 
\bea\label{expansion}
f(x,y) = \frac{(x-y)^2}{x+y}
\left(1 +\xi_2 \left(\frac{x-y}{x+y}\right)^2 +\cdots\right).
\eea
As mentioned earlier, the higher-derivative expansion takes the form of a power series
in $N^{-2/3}$ or, equivalently, in inverse powers of the square root $\sqrt{V_4}$ of the
discrete four-volume (the number of four-simplices). 
It is our task to determine from the data whether or not such 
subleading terms are present in the effective action.
However, let us emphasize that the presence of higher-derivative corrections
at the regularized level, like those  
associated with the coefficients $\xi_1$ and $\xi_2$ in (\ref{poteff}) and
(\ref{expansion}), does not 
imply that such terms necessarily survive in the continuum limit. 
For example, if in the simulations we observed a coefficient 
$\xi_2$ which was independent of the four-volume $V_4$
of the universe, we could convert the dependence on $1/\sqrt{V_4}$ 
of the corresponding term in the action into
a dependence on the lattice spacings $a_t$ and $a_s$ 
in the time and spatial directions according to
\beq\label{volume}
V_4^{cont} \propto a_t a_s^3V_4,
\eeq 
or $a^4 V_4$ in shorthand notation\footnote{The notation ``$a$" for a generic
lattice spacing should not be confused with the same notation for the scale
factor $a(t)$.}.
In this case the term $(x-y)^2/(x+y)^2$ in (\ref{expansion}) would be proportional 
to $a^2$ and simply drop out in a standard scaling 
limit where we keep $V_4^{cont}$ constant while taking $V_4$ 
to infinity and the lattice spacings $a_t$ and $a_s$ to zero. 
Likewise, if we wanted to add such a higher-derivative term by hand 
already in the bare
action, such that it survived in the classical limit, we would 
need to assume an appropriate nontrivial 
lattice dependence of the coefficient $\xi_2$ in front of the 
discretized higher-derivative term. 
However, we will not follow this latter route in the present work. This
does not imply that no higher-curvature terms may appear in the effective action,
it only means that no tunable coupling constant is associated with such terms,
and that the coefficient $\xi_2$ is determined purely from the entropy of microstates.

In the region where we can presently perform reliable 
measurements, $\xi_2$ does not display any significant scale dependence. 
However, it could in principle
pick up such a dependence when we move closer to 
a potential ultraviolet fixed point by
changing the bare coupling constants, along the lines described in \cite{bigs4}.
In this way, in the vicinity of a nontrivial ultraviolet fixed point
(in principle infinitely many) higher-derivative terms can play a role.
In view of this situation our task is to first identify potential higher-derivative 
terms in the effective action, and then study 
the scaling behaviour of the associated coupling constants.

Let us determine the effect of the presence of a $\xi_2$-dependent
term on the fluctuation matrix $S_{n,n'}$.
We can use the fact that our tentative discretized action
should reproduce the ``observed'' $N_{cl}(n)$ of \rf{newsphere} 
as an extremum. 
Extremality is satisfied if
\bea
V(N(n)) = \frac{9}{2} \left(-\frac{A_t^2}{\sqrt{ V_4 }} N(n)+
\left(\frac{3}{4}\right)^{2/3}A_t^{8/3} N^{1/3}(n) + 
{\cal O}(N^{-1/3}(n))\right).
\label{volexpand}
\eea
To this order the result does not depend on 
the parameters $\xi_1$ and $\xi_2$. Note that $A_t$ 
can be fully determined from the data for $N_{cl}(n)$. 
Furthermore, the additional term proportional to $\eps$ 
gives no contribution to the classical solution, since
the bare cosmological constant $\kappa_4$ was chosen
to achieve exactly that, as explained at the end of Sec.\ \ref{sec2}. 

We can now use (\ref{seff}) to derive a prediction for the 
inverse covariance matrix, namely, 
the matrix $S_{n,n'}$ of second derivatives at $N(n)=N_{cl}(n)$
(see \rf{derivative}), provided that
the fluctuations around $N_{cl}(n)$ can be approximated 
by working up to second order in the $\del N(n)$ 
(see \cite{bigs4} for a detailed discussion).
It is a simple exercise to show that the matrix elements 
(after again eliminating a constant
$2\eps$-shift) satisfy 
\bea\label{j8}
S_{n,n} +\frac{N_{cl}(n+1)}{N_{cl}(n)}S_{n,n+1}+
\frac{N_{cl}(n-1)}{N_{cl}(n)}S_{n,n-1} = \frac{1}{\Gamma} V''(n) 
\eea
where
\bea\label{j9}
V''(n)\equiv V''(N_{cl}(n))=
-\left(\frac{3}{4}\right)^{2/3}A_t^{8/3}N_{cl}^{-5/3}(n) + 
{\cal O}(N^{-7/3}(n)).
\eea
As long as we keep only the leading term in \rf{j9}, 
the right-hand side of \rf{j8} is independent of $\xi_2$.
However, the elements $S_{n,n}$ and $S_{n,n\pm 1}$ can (and do) 
depend on $\xi_2$ at 
this order, provided these contributions cancel on the 
left-hand side of \rf{j8}. 
\begin{figure}
\centering
\includegraphics[width=0.75\textwidth]{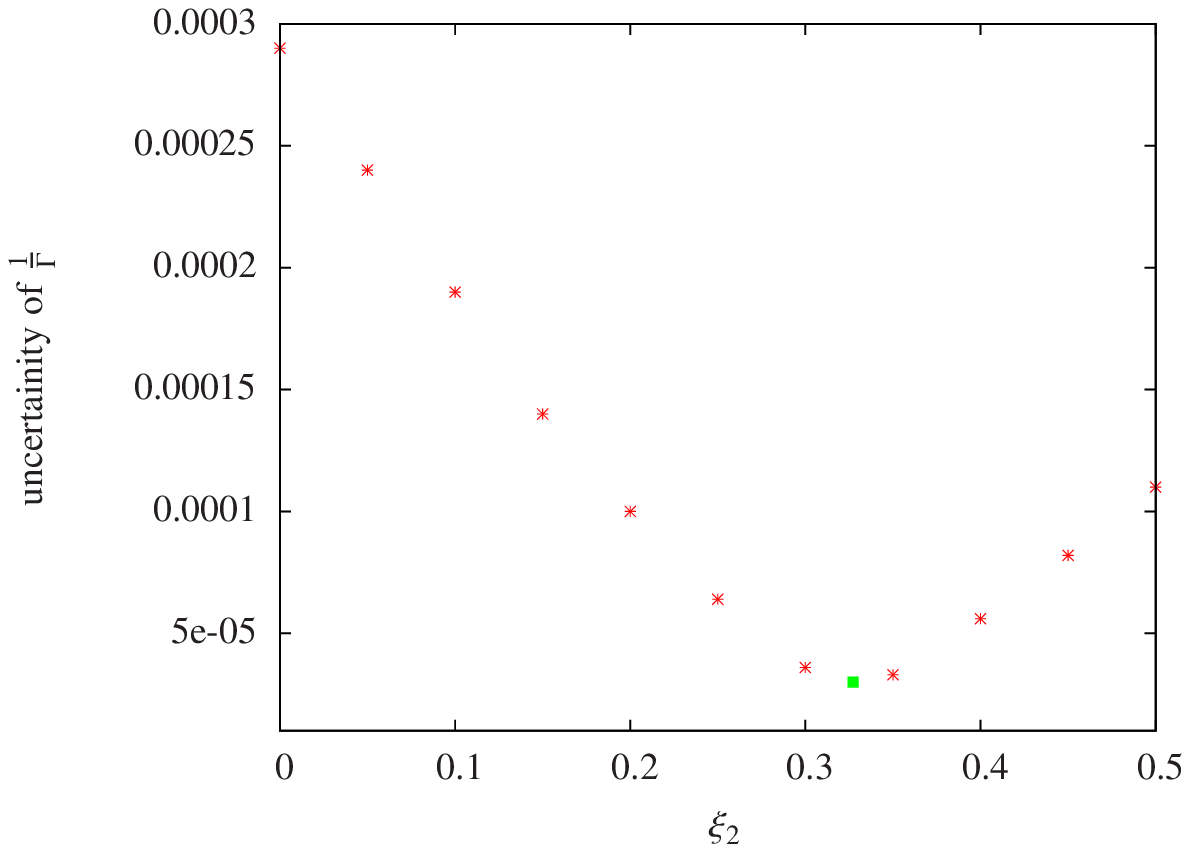}
\includegraphics[width=0.75\textwidth]{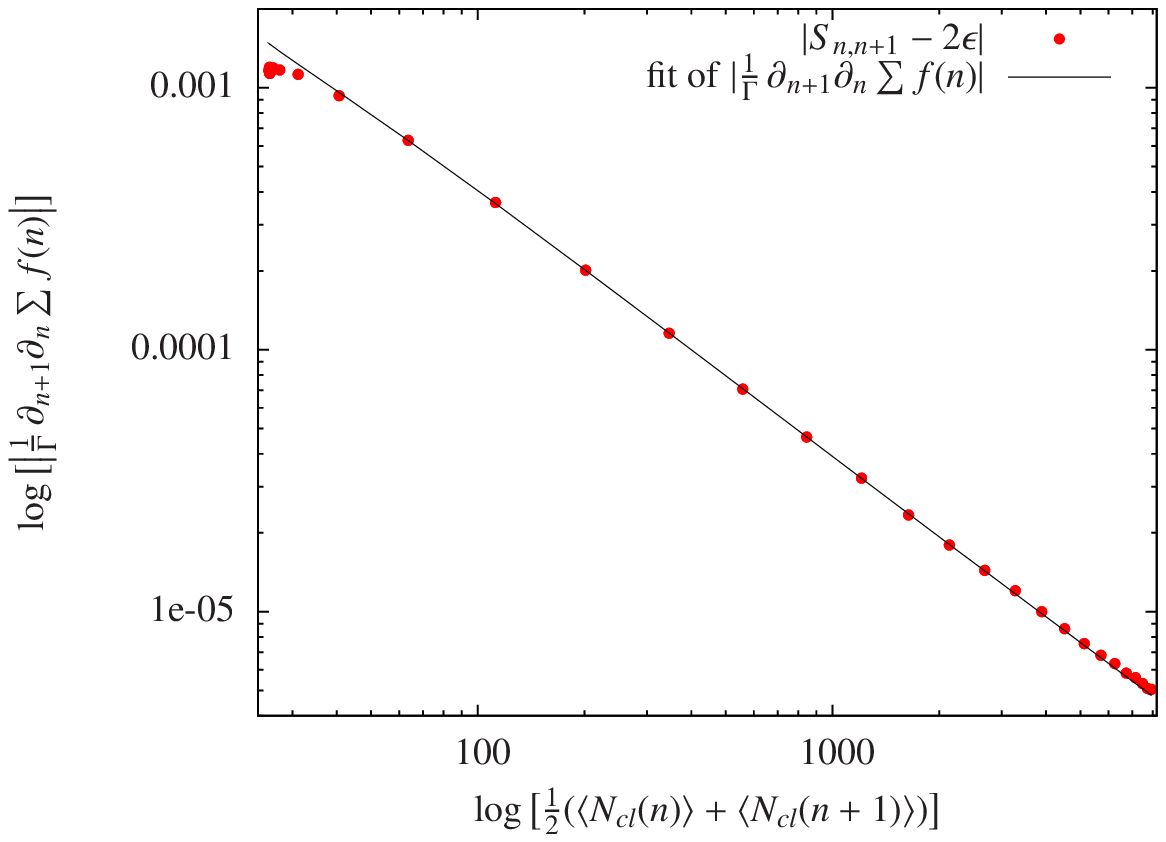}
\caption{Determining the parameter $\xi_2$ for the standard choice of couplings. 
Top: finding the optimal value of $\xi_2$ by minimizing the error of $1/\Gamma$
in the fit for the subdiagonal $S_{n,n+1}$ (best fit denoted by the green square). 
Bottom: the best fit in a log-log plot of $|S_{n,n+1}-2\epsilon |$ versus the
averaged three-volume $\langle N_{cl}(n)+N_{cl}(n+1)\rangle /2$.}
\label{Fig7}
\end{figure}
This is illustrated by the expression for $S_{n,n\pm1}$, which 
after subtraction of a $2\eps$-contribution is given by
\bea\label{jurek1}
S_{n, n\pm1} = -\frac{1}{\Gamma}
\frac{8 N_{cl}(n)N_{cl}(n\pm 1)}{(N_{cl}(n)+N_{cl}(n\pm 1))^3}
\left(1 + 6\xi_2\left(\frac{N_{cl}(n\pm1)-N_{cl}(n)}{N_{cl}(n\pm1)+
N_{cl}(n)}\right)^2+\cdots\right).
\eea
In Fig.\ \ref{Fig7} we compare this prediction with the data 
for the standard choice $(\kappa_0,\Delta)=(2.2,0.6)$ of bare constants.
Notice that the term proportional to $\xi_2$ is 
${\cal O}(V_4^{-1/2})$ with respect to the leading term, 
exactly as in the  ``potential" $V''(n)$.
Fig.\ \ref{Fig7} shows the results of the analysis 
including the leading correction to the kinetic term. 
The best fit is obtained for $\xi_2 = 0.327\pm 0.010$.
This concludes our discussion of how to search for and analyze higher-order
correction terms to the effective action. For such terms to have a continuum
interpretation as a curvature-squared contribution like in the
continuum action (\ref{r2action}), we need to show that a whole bunch
of coupling constants like $\xi_2$ exhibit the correct scaling behaviour as 
we follow their flow when we approach the critical phase transition lines.
Our present Monte Carlo algorithm still needs
improvement to deal with the critical slowdown appearing close to 
the B-C transition line, which in this context
has our primary interest \cite{bigs4,horava-simu}.

In the appendix we repeat the above analysis for the system with
subdivided time steps, analogous to what we did in Sec.\ \ref{sec4} for
the volume distribution. This turns out to be self-consistent, in the sense
that integrating out the substructure at the discretized level produces a
result compatible with what was derived for integer time steps
in the present section.

\section{Summary and discussion}\label{sec6}

In the work presented above, we have analyzed several properties of the 
semiclassical limit of CDT quantum gravity, a limit which is inherently
nonperturbative in nature. We used a modified simulation method
which replaces the fixing of the total four-volume to a specific value $V_4$
by a Gaussian distribution peaked at $V_4$. In this way one avoids the
presence of a zero-mode of the volume fluctuations, whose removal 
had hampered the extraction of short-distance continuum physics in 
previous work.

After reconfirming our previous result for the de Sitter volume profile we measured
the probability distributions ${\cal P}_n(N)$ for finding a three-volume
$N$ at a fixed, given time step $n$. Well inside the universe, we found them 
to be Gaussian around the average, confirming their semiclassical nature.
Moving toward the `stalk', where $N$ becomes small, 
the probability distributions become asymmetric and  
display discretization artifacts, 
in the sense that the data in each ${\cal P}_n(N)$ can be seen to 
separate into three distinct families. These
effects are confined to very small length scales, but -- depending on
the type of observable under study -- they can potentially interfere 
with the nonperturbative {\it physical} short-distance effects we are
keen to explore, for example, along the lines suggested in \cite{reuter}.

The volume profile is an example of an observable which is not particularly
sensitive to these discretization artifacts, even down to short scales.
We found further evidence for this remarkable robustness when refining
our spatial slicing, effectively doubling the number of slices in a given time
interval. This was done by identifying the connected (on the dual lattice)
layers of (3,2)-simplices in between successive (4,1)-layers with slices of
fixed half-integer time. Determining the volume profile for the (3,2)-layers
alone, we found that after rescaling their volume by a constant $\rho$, 
the resulting profile could be combined with the previous data for the 
(4,1)-slices to yield a single, universal volume distribution.   
This analysis could be extended further by splitting each time interval
$[n,n+1]$ not just into two but into three, yielding a similar outcome.
Our results show that the macroscopic geometry associated with
the semiclassical limit of CDT quantum gravity is not closely linked
to the microscopic piecewise linear geometry of the building blocks,
in line with the nonperturbative character of this limit.
Our attempt to learn more about the nature of the A-C phase transition
by studying the scaling behaviour of the constant $\rho$ as a function
of the bare parameters $\kappa_0$ and $\Delta$ yielded at this stage
inconclusive results.

Our new method of volume-fixing also enabled us to make new precision
measurements of the effective action for the three-volume fluctuations $\delta N(n)$
of the universe around its semiclassical limit. Specifically, we investigated
correction terms to the action -- obtained by inverting the covariance matrix
$\langle \delta N(n) \delta N(n')\rangle$ -- which are of higher order in
powers of $N^{-2/3}(n)$ and are associated with curvature-squared terms in
the continuum. We observed nonvanishing correction terms of this type in the
discretized action, while others, expected from a comparison with the continuum,
gave little or no contribution. This neither proves nor disproves the presence
of higher-order curvature terms, since we found that their coefficients in the
discrete effective action must have a specific, nontrivial scaling
behaviour (as a function of the UV cut-off $a$) in order to survive in the limit
$a\rightarrow 0$. We undertook a careful, quantitative analysis of one of
these coefficients, but did not find evidence for the required scaling, at least
not for the range of coupling constants considered. The implication is that 
there is currently no evidence for the presence of higher-order curvature
in CDT quantum gravity, although this issue needs to be re-examined 
once we have improved our algorithms to penetrate to shorter distance
scales by performing the simulations closer to the putative UV fixed points
in the phase diagram.

\vspace{1cm}

\noindent {\bf Acknowledgements.} This work is partly supported by 
the International PhD Projects Programme of the Foundation for Polish 
Science within the European Regional Development Fund of the European 
Union, agreement no. MPD/2009/6.
AG and JJ ackonowledge a partial support by the 
Polish Ministry of Science grant N~N202~229137 (2009-2012). 

\vspace{1cm}

\section*{Appendix}

In this appendix, we repeat the analysis of Sec.\ \ref{sec5} for the
system where we have subdivided all time steps into two, as already
described in Sec.\ \ref{sec4}. That is, we associate
constant half-(odd-)integer times to the layers made up of $(3,2)$- and 
$(2,3)$-simplices. 
We want to investigate whether this finer-grained system
can also be described by a discretized  
semiclassical effective action. If so, the action we have already 
derived, based on the data from integer-valued constant-time slices alone
should be obtainable from the new effective action
by integrating out the $(3,2)$- and $(2,3)$-``degrees of freedom''.
In this way the system described by the integer-time slicing 
may be understood as arising from a ``Kadanoff blocking'' in time of the 
larger, finer-grained system. 

After measuring the covariance matrix
of ``three-volume'' fluctuations\footnote{We put ``three-volume"
in quotation marks because -- unlike the number of 
(4,1)-simplices -- the number of (3,2)- and (2,3)-simplices
in a slab between times $n$ and $n+1$ does not have a direct
interpretation in terms of a three-volume.} for all integer and 
half-integer times, we invert this matrix to obtain
the effective action, up to quadratic terms in the fluctuations.  
The large inverted covariance matrix can be decomposed 
into blocks describing the (4,1)-system (at integer time),
the (3,2)+(2,3)-system (at half-integer times) and off-diagonal 
blocks describing interactions between the two.
As expected, we observe a constant shift by $2\eps$ in the 
(4,1)-block only, which we subtract from the entries of the
fluctuation matrix. 

The functional form of the matrix entries 
resembles that found in Sec.\ \ref{sec5}, but with some modifications.
Before proceeding further, let us rescale the number of (3,2)+(2,3)-vertices
with the factor $\rho$ determined in Sec.\ \ref{sec4} to obtain
the number $N(n+1/2)$ which is part of the universal volume
distribution $N(n)$, valid for both integer and half-integer $n$.
The measured matrix structure is consistent with an
effective action of the form
\bea
S_\textit{eff}^A &=& \frac{1}{\Gamma_1} 
\sum_n \Big( g(N(n+1/2),N(n))+g(N(n-1/2),N(n))\Big)\\ \nonumber
&-&\frac{1}{\Gamma_2} 
\sum_n \Big( f(N(n),N(n+1))-V_1(N(n+1/2))-V_2(N(n))\Big)\\ \nonumber
&+&\eps\left(\sum_n N(n) - V_4\right)^2 ,
\eea
where
\bea
g(x,y)&=&\frac{(x-y)^2}{x+y}
\left(1 + \chi_1\left(\frac{x-y}{x+y}\right)+
\chi_2\left(\frac{x-y}{x+y}\right)^2+\cdots\right),\\ \nonumber
f(x,y)&=& \frac{(x-y)^2}{x+y}
\left(1 +\chi'_2 \left(\frac{x-y}{x+y}\right)^2 + \cdots\right),
\eea
and
\begin{align}
V_{1}(N(n+1/2))&=\mu^{(1)}N(n+1/2)^{1/3}-\lambda_\textit{eff}^{(1)}N(n+1/2), \\ 
V_{2}(N(n))&=-\mu^{(2)}N(n)^{1/3}+\lambda_\textit{eff}^{(2)}N(n). \nonumber
\end{align}
We observe that the ``kinetic'', finite-difference terms 
couple not only neighbouring layers 
(in this case $n$ and $n+1/2$), but also directly
the (4,1)-layers at $n$ to those at $n\pm 1$, in the latter case with 
a {\it negative} sign. We have checked that this feature persists
when we perform a further ``time refinement" by associating the
(3,2)- and (2,3)-simplices with distinct spatial slices, as was done
in Sec.\ \ref{sec4} above.  
There are further unexpected minus-signs in
the potential $V_2$, and a new correction term linear in $(x-y)$ in the
kinetic term $g(x,y)$. We take this as an indication that -- unlike what
happened in our earlier analysis of the volume distribution 
itself -- we run into short-distance lattice artifacts for measurements
of the volume fluctuations when considering the subdivided slicings.

Demanding that integrating over the (3,2)-fluctuations 
should give back the (4,1)-effective action leads to
explicit relations among $\Gamma$, $\Gamma_1$,
$\Gamma_2$, as well as among $\mu$, $\mu^{(1)}$, $\mu^{(2)}$ 
and $\lambda_\textit{eff}$, $\lambda_\textit{eff}^{(1)}$, 
$\lambda_\textit{eff}^{(2)}$. A simple, but tedious
computation yields
\bea
\frac{1}{\Gamma}=\frac{1}{2\Gamma_1}-\frac{1}{\Gamma_2}
\label{ccs1}
\eea
and
\bea
\label{ccs2}
\mu&=&\big(-\mu^{(1)}+\rho^{-1/3}\mu^{(2)}\big)
\frac{2\Gamma_{1}\Gamma_{2}}{\Gamma_{2}-2\Gamma_{1}}\Gamma_{2}, \\[1ex]
\lambda_\textit{eff}&=&\big(-\lambda_\textit{eff}^{(1)}+\rho^{-1}\lambda_\textit{eff}^{(2)}\big)
\frac{2\Gamma_{1}\Gamma_{2}}{\Gamma_{2}-2\Gamma_{1}}\Gamma_{2}, \nonumber
\eea
as well as relations between $\chi_i$, $\chi'_i$ and $\xi_i$. 

We have checked numerically that 
starting out with the larger covariance matrix of the subdivided system
and integrating out the (3,2)-fluctuations reproduces within measuring
accuracy the covariance matrix determined earlier by considering
integer times only. 
The results at the point $(\kappa_0,\Delta)=(2.2,0.6)$ in phase C 
are presented in the table below, which compares the directly measured
\begin{center}
\begin{tabular}{|c|c|c|}
\hline
parameter & direct  & integrated  \\ 
& (from (4,1)) & (from (4,1) and (3,2)) \\
\hline
$\Gamma$ & $23 \pm 1$ & $26 \pm 2$ \\ 
\hline
$\mu$ &  $13.9 \pm 0.7$ & $11.3 \pm 0.7$ \\
\hline
$\lambda_\textit{eff}$ & $\,\,\,0.027 \pm 0.003$ & $\,\,\,0.027 \pm 0.003$ \\
\hline
\end{tabular}
\end{center}
parameters of the (4,1)-action (\ref{seff}) with their
counterparts obtained from measuring the fine-grained system and
using formulae (\ref{ccs1}) and (\ref{ccs2}).
In the fits we have neglect higher-order corrections,
and the errors are estimated as statistical errors of the fits.
In view of this the agreement between the two sets of values appears quite
satisfactory.

\begin{figure}
\centering
\includegraphics[width=0.45\textwidth]{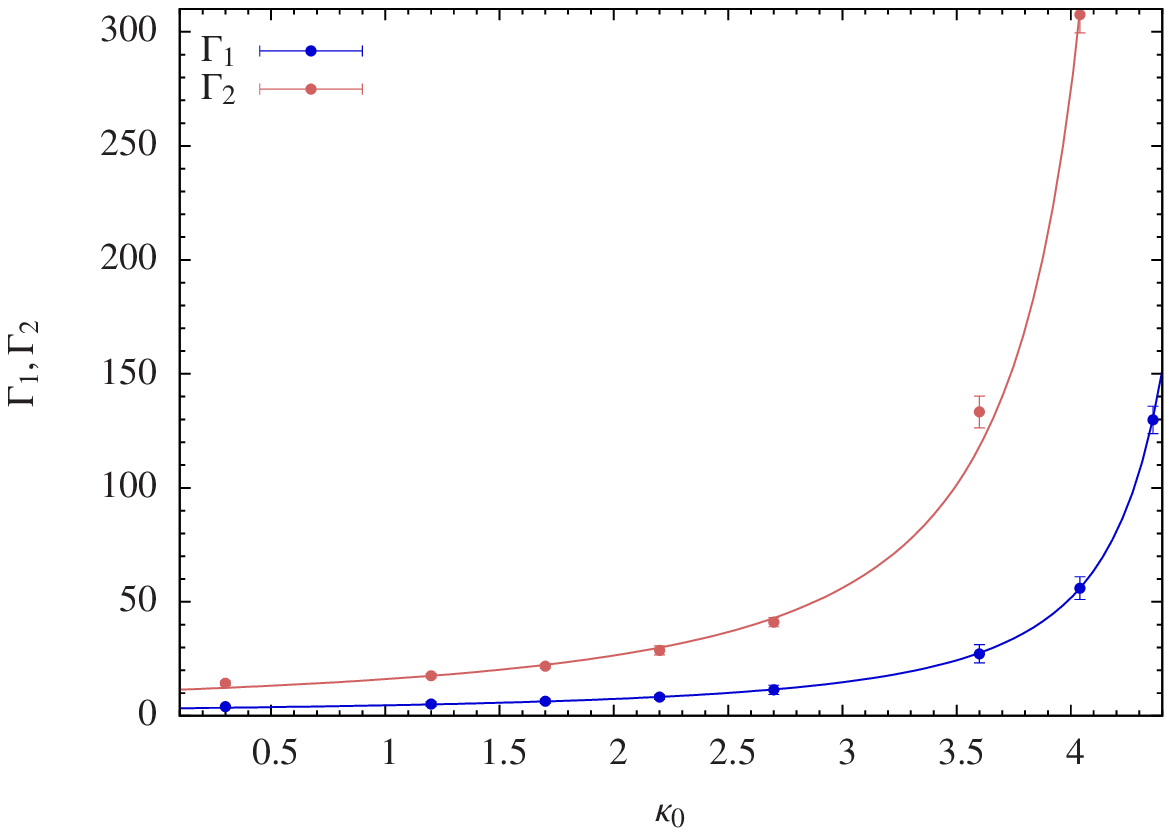}
\includegraphics[width=0.45\textwidth]{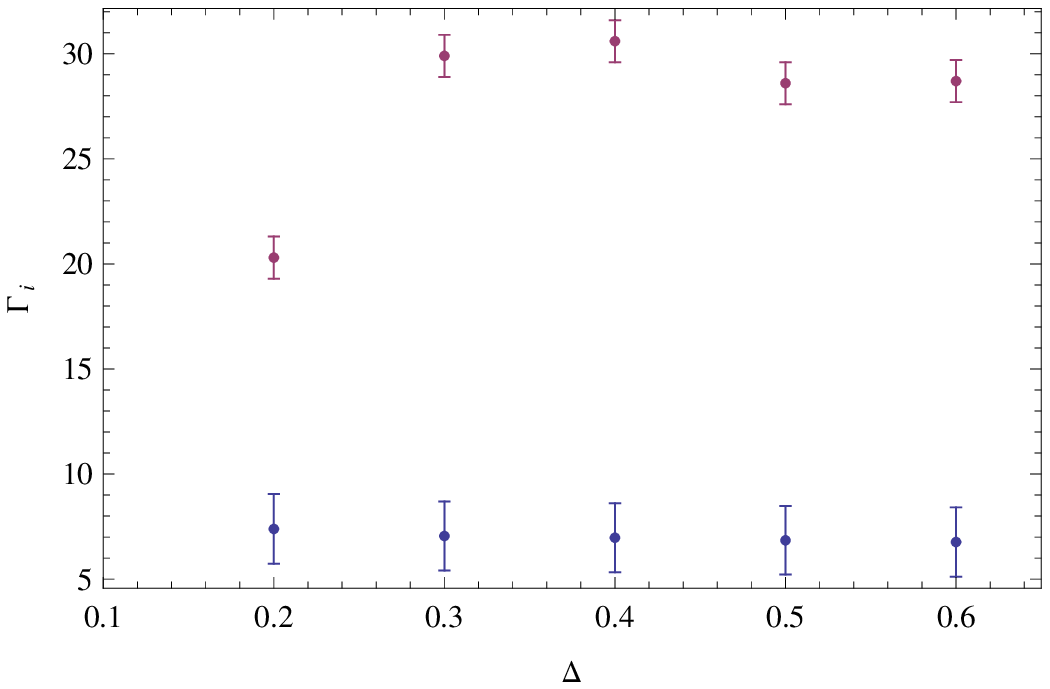}
\caption{Dependence of the parameters $\Gamma_1$ (lower curves) 
and $\Gamma_2$ (upper curves) 
on $\kappa_0$ (left figure) and $\Delta$ (right figure)
for selected points on the phase diagram.}
\label{figgamma}
\end{figure}

We have also examined the dependence of the 
action parameters $\Gamma_i$ on the bare coupling constants, 
using the same values as in the analogous investigation of the parameter
$\rho$ in Sec.\ \ref{sec4}. For constant $\Delta$ we observe a 
marked dependence on $\kappa_0$ (see Fig.\ \ref{figgamma}), in the
sense that both $\Gamma_1$ and $\Gamma_2$ seem to diverge when 
approaching the critical line for the A-C phase transition. 
For fixed $\kappa_0$ we observe a nontrivial behaviour of $\Gamma_2$,
which may be corroborating evidence that the
semiclassical scaling function \rf{sphere}
ceases to give a good description of the three-volume distribution
around $\Delta \approx 0.3$.

\end{document}